\documentclass[twocolumn]{aastex631}

\newcommand{\dd}{submicrometer-sized grains}

\newcommand{\simgreat} {\mathbin{\lower 3pt\hbox{$\rlap{\raise
				5pt\hbox{$\char'076$}}\mathchar"7218$}}}
\newcommand{\simless}{\mathbin{\lower 3pt\hbox {$\rlap{\raise
				5pt\hbox{$\char'074$}}\mathchar"7218$}}}
\shorttitle{Luminosity distance and extinction by submicrometer-sized grains}
\shortauthors{R.~Siebenmorgen et al.}

\begin{document}

\title{Luminosity distance and extinction by \dd}

\author{R.~Siebenmorgen}
\affiliation{European Southern Observatory,\\
	Karl-Schwarzschild-Str. 2, 85748 Garching, Germany}
\author[0000-0002-3245-4272]{Frank Heymann}
	\affiliation{German Aerospace Center \\
		Institute for solar terrestial physics \\
		Kalkhorstweg 53, 17235, Neustrelitz, Germany}
\author{R.~Chini}
\affiliation{Nicolaus Copernicus Astronomical Center of the Polish Academy of
	Sciences, \\
	Bartycka 18, 00-716 Warsaw, Poland; }
\affiliation{Ruhr University Bochum, Faculty of Physics and Astronomy,\\ Astronomical Institute (AIRUB),\\ 
44780 Bochum, Germany;}
\affiliation{Universidad Catolica del Norte, Instituto de
	Astronomia,\\
	Avenida Angamos 0610, Antofagasta, Chile}



\begin{abstract}
\noindent The distance to the stars is a fundamental parameter, which
is determined via two primary methods – parallax and luminosity. While
the parallax is a direct trigonometric method, the luminosity distance
is usually influenced by interstellar extinction. As long as the
optical properties of dust grains are wavelength-dependent this
contamination can be corrected. \replaced{With increasing grain size,
  however, the extinction properties become gray, i.e. they contribute
  at wavelengths but can no longer be detected by photometry.}
{However, as the grain size increases, the extinction properties
  become gray, meaning these particles contribute by a constant at
  wavelengths $\simless \,1\mu$m, making them undetectable by
  photometry in the optical.}  In this study we compare the
parallactic and luminosity distances of {  {a pristine sample of}}
33 well-known early-type stars with non-peculiar reddening curves and
find that the luminosity distance overestimates the parallactic
distance in 80\% of the cases. This discrepancy can \deleted{only} be
removed when incorporating a population of large, submicrometer-sized
dust grains in a dust model that provides gray extinction which
diminishes the luminosity distance accordingly.
\end{abstract}


\section{Introduction} \label{sec:intro}

The distance $D$ of astronomical objects is a fundamental { 
  {quantity}}, prompting the exploration of various estimation
techniques. The parallax $\pi$, based on geometric principles,
provides the most straightforward approach but is limited to nearby
stars. The luminosity distance $D_{{L}}$ enables the assessment of
distances of many kpc by measuring the apparent stellar brightness and
comparing it with the absolute brightness. However, the apparent
brightness of stars is usually diminished by an unknown amount of
dust.

Traditionally, the extinction is determined from the apparent color of
stars in the visible range (e.g. $B - V$), comparing it with the
unreddened color $(B - V)_0$, and multiplying the difference $E(B -
V)$ with the ratio \deleted{$R_V$, } of total-to-selective
extinction\added{, $R_V$}. The average value of $R_V$ in the Milky Way
is $\sim 3.1$ \citep{45}, with extreme values from 2.1 in clouds of
high galactic latitude to 5.6 in dense molecular clouds. It should be
noted that $R_V$ corresponds to the extinction at infinite wavelengths
and is extrapolated from measurements at near-infrared (NIR)
wavelengths. The reddening longward of $2.2,\mu$m is more difficult to
establish because contamination by either dust or any other emission
components of early-type stars might \replaced{come into play}{play an
  important role} \citep{S18, Deng22}.

By introducing the photometric equation nearly 100 years ago it was
speculated that an additional non-selective or gray extinction term in
the form of very large grains – at that time called 'meteoritic'
bodies – might exist, which increases the extinction term in
Eq.~\ref{Eq.1} by a constant offset but which was arbitrarily set to
zero \citep{8}

\begin{equation} \label{Eq.1}
	A_V = V - M_V - 5 \log D_{{L}} + 5 \ . 		
\end{equation}

where $M_{V}$ is the absolute magnitude of the star. Meanwhile,
submicrometer-sized particles have been found in various environments
(Sect.~\ref{dust.sec}) providing significant wavelength-independent
extinction from the far-UV to the NIR. As a consequence, $A_V$
determinations from optical reddening must underestimate the total
extinction and thus lead to larger distances. For a few stars, it was
hypothesized \citep{9} that incorporating an additional component of
large dust grains could reconcile the problem of missing extinction
and thus inaccurate distance estimates; however, this remained
unverified due to the lack of a physical model.

In the following we use distances $D_{\rm {Gaia}}$, obtained from the
zero-point corrected parallax following \cite{Lindegren21} and compare them with the luminosity distances $D_{{L}}$
     {  {for a pristine sample of}} 33 well-known OB stars.  For
     many stars, our investigation uncovers an inconsistency between
     both methods; its detection required the unprecedented resolution
     of the Gaia data release three \citep{DR3}. To reconcile the
     discrepancy in distance, we consider a contribution of gray
     extinction (Eq.~\ref{Eq.1}) due to \dd. In the ISM a {  {gray
         component of micrometer-sized grains was introduced by
         \cite{Mathis77} and recently by}}  \cite{WL15a, WL15b}
     to account for the observed infrared extinction. \added{Finally,}
     we calibrate the contribution of \replaced{micrometer-sized}{\dd}
     to the total extinction with the distance suggested by Gaia.

     \section{The sample }

The inclusion of an additional extinction term necessitates validation
through a dust model. For such a test a well-selected sample of
reddening curves is crucial.  The stars shall have precise photometric
measurements, accurate distance determinations, rigorous spectral type
classifications and luminosity class identifications, allowing
reasonable estimates of the absolute brightness $M_{V}$, as well as
high-quality reddening curves spanning the entire range from the Lyman
limit to infinitely long wavelengths.

In that wavelength range, 820 reddening curves have been published by
\cite{V04,FM07,G09}, which suffer from various systematic
uncertainties. To obtain a high-quality sample, \cite{1} inspected and
merged a sample of 186 stars, mainly observed with the Ultraviolet and
Visual Echelle Spectrograph (UVES; \citep{UVES}). From this sample,
stars with composite spectra in the IUE/FUSE apertures arising from
multiple bright stellar systems were excluded. Likewise, stars were
omitted with a photometric variability between space
\citep{Kharchenko09, DR2, DR3} and ground-based observations
\citep{V04} of more than $\sigma(V) = 30$\,mmag, $\sigma(B - V) =
30$\,mmag, and $\sigma(G) = 11$\,mmag, respectively. Further, stars
that show inconsistent parallaxes when comparing data releases two and
three from Gaia \citep{DR2,DR3} were rejected.

The spectral type and luminosity class (SpL) estimates of these stars
are important for the determination of the reddening curve and when
utilizing $M_{V}$ for distance estimates (Eq.~\ref{Eq.1}). The SpL
were identified by fitting the \cite{GC14} library to UVES spectra
with undetected signatures of binarity or circumstellar line
profiles. This provided a precision of half a sub-type/class, which
matches the accuracy reached by other studies \citep{Kyritsis22,
  Liu19}. For the O-stars, it was also confirmed that the SpL
classification agrees with that provided by the Galactic O-star survey
\citep{Sota14}. Stars whose classification does not confirm the one
used in the reddening curve determination were removed.

{  {We apply even stricter criteria to the Gaia data. Only stars
    with a renormalized unit weight error (RUWE) below 1.1 are
    included, ensuring that a single-star model accurately fits the
    astrometric solution \citep{Luri18}. Additionally, we re-compute
    the $G$-magnitude-dependent error in parallax $\sigma(\pi,G)$
    (Eq.\ref{Eq.2err}) following \cite{MA22}, and only consider stars
    with a parallax precision of $\pi / \sigma(\pi,G) > 10$.  The
    simple inverse of the DR3 catalogue parallax typically agrees with
    $D_{\rm {Gaia}}$ within $1.6 \pm 1.8$\,\%. Since parallactic
    distances inherently depend on priors, we verified that our
    distance estimate $D_{\rm {Gaia}}$ aligns with other probabilistic
    distance estimates within 1–2\% \citep{BailerJones21}. From the
    fifty stars that satisfy the criteria for the reddening curve, an
    additional 17 stars (34\%) were rejected for not meeting the
    stricter Gaia data requirements. Ultimately, using the most robust
    sample of reddening curves, spectral classification, and
    parallaxes currently available, we identified a pristine sample of
    33 sightlines toward prominent OB stars within a distance of
    2.5\,kpc. }}


\section{The distance discrepancy \label {disc.sec}}
\begin{figure} [!htb]
	\begin{center}
		\includegraphics[width=9cm,clip=true,trim=4.2cm 5.5cm 3.cm 5.6cm]{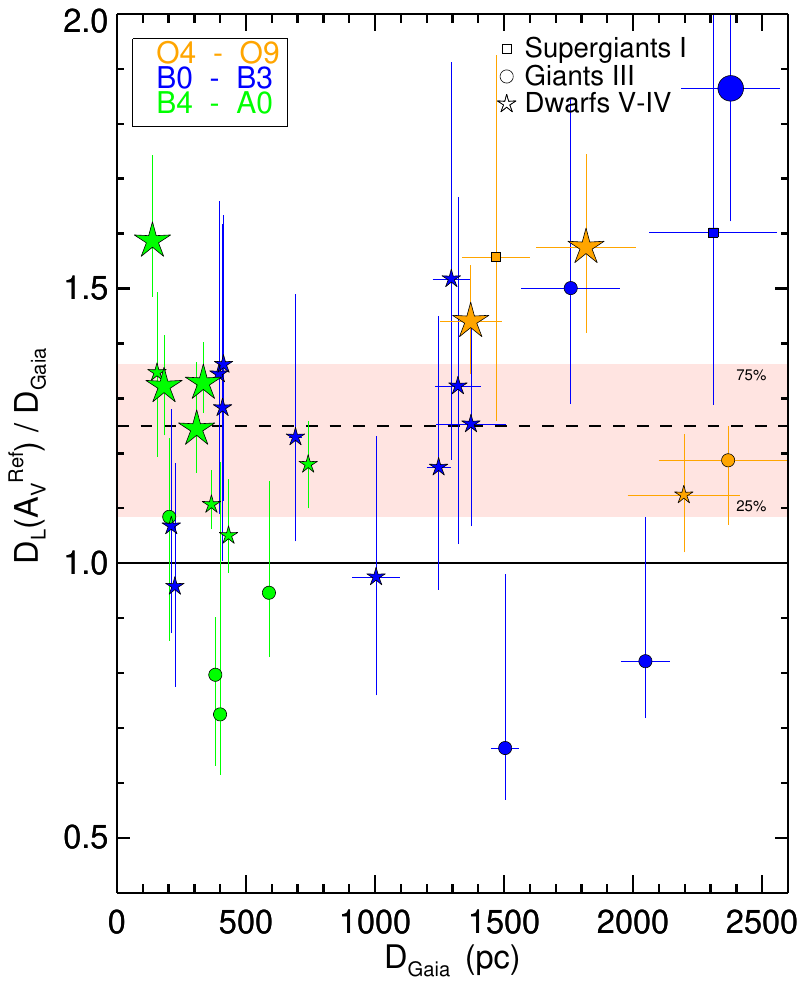}
	\end{center}
	\caption{The distance ratios $D_{{L}}(A^{\rm {Ref}}_V)/D_{\rm {Gaia}}$
		vs. $D_{\rm {Gaia}}$ for our sample of 33 stars using visual
		extinction $A_V^{\rm {Ref}}$ estimates obtained by extrapolation of
		the reddening curves \citep{V04, FM07, G09}. Different symbols and colors
		are used to visually represent the various spectral types and
		luminosity classes as labelled. Stars with a deviation in the
		distance ratio from unity that are below $3\sigma$ in $M_{V}$ are
		shown by small symbols. The shaded area depicts the top and bottom
		quartiles of the distribution, with the mean indicated by a dashed
		line.}
	\label{DL2DGorg.pdf}
\end{figure}

The luminosity distance $D_{{L}}(A^{\rm {Ref}}_V)$ was computed
using Hipparcos photometry \citep{Kharchenko09}; spectral types and
luminosity classes are from UVES \citep{1}. Absolute magnitudes $M_V$
were calculated from our SpL and the conversion tables \citep{40,
  41}. The visual extinction $A^{\rm {Ref}}_V$ was obtained by
extrapolation of the reddening curves \citep{V04, FM07, G09}. The
errors in $D_{{L}}(A^{\rm {Ref}}_V)$ primarily arise from
systematic uncertainties. We apply the conservative error estimate
$\sigma(M_V)$ in $M_V$ following \cite{40}. The $1 \sigma$ scatter
between $M_V$ of the spectral type and the adjacent $\pm 0.5$ subtype
is denoted as $\sigma_{\rm {SpT}}$, the $1 \sigma$ scatter between
$M_V$ of the luminosity class $\pm 1$ luminosity class is denoted as
$\sigma_{\rm{LC}}$.
%
\begin{table*}[!htb]
	\begin{center}  
		\caption {Data supporting Fig.~\ref{DL2DGorg.pdf} and Fig.~\ref{DL2DGcor.pdf}.} \label{Tab1}
		\begin{tabular}{c l c c c c c c r r r}
			\hline
			\hline
			1 & 2&  3&  4& 5&  6& 7& 8&  9&  10& 11 \\
			\hline  
			Star & Ref & SpL & $M_{V}$ &$V$  &    $E(B-V)$ & $A^{\rm {Ref}}_{V}$ & $A_{V}$ & $D_{{L}}(A^{\rm {Ref}}_{V})$ & $D_{\rm {Gaia}}$ & $D_{{L}}(A_{V})$ \\
			&     &     & (mag)& (mag)   & (mag)           & (mag)        & (mag)   & (pc)           & (pc)  & (pc)  \\
			\hline
			& &  &  & &  & & &  &  &  \\
			\smallskip
			\smallskip
			HD~027778 &   G &          B3V  & $-1.52   _{-0.44} ^{+ 0.39 }$&  6.33 &  0.39  &  1.09  &$ 1.23\pm  0.42  $&$   225    _{ -41 }^{+   45}$ & $  210 \pm    3$ &  210 \\
			\smallskip
			HD~037903 &   G &          B2V  & $-2.28   _{-0.46} ^{+ 0.46 }$&  7.84 &  0.36  &  1.49  &$ 2.13\pm  0.46  $&$   531    _{-100 }^{+  124}$ & $  395 \pm   10$ &  395 \\
			\smallskip
			HD~038023 &   F &          B3V  & $-1.52   _{-0.44} ^{+ 0.39 }$&  8.87 &  0.52  &  1.64  &$ 2.31\pm  0.42  $&$   562    _{-103 }^{+  112}$ & $  413 \pm    7$ &  413 \\
			\smallskip
			HD~046223 &   V &          O4V  & $-5.65   _{-0.13} ^{+ 0.13 }$&  7.31 &  0.54  &  1.48  &$ 2.27\pm  0.23  $&$  1974    _{-118 }^{+  126}$ & $ 1371 \pm  120$ & 1370 \\
			\smallskip
			HD~054439 &   F &          B1V  & $-3.04   _{-0.53} ^{+ 0.50 }$&  7.71 &  0.31  &  0.80  &$ 0.75\pm  0.55  $&$   979    _{-212 }^{+  254}$ & $ 1004 \pm   92$ & 1004 \\
			\smallskip
			HD~062542 &   G &          B5V  & $-1.21   _{-0.09} ^{+ 0.12 }$&  7.99 &  0.37  &  1.16  &$ 1.38\pm  0.11  $&$   405    _{ -16 }^{+   23}$ & $  366 \pm    5$ &  366 \\
			\smallskip
			HD~070614 &   F &        B3III  & $-2.85   _{-0.33} ^{+ 0.85 }$&  9.29 &  0.68  &  2.14  &$ 1.93\pm  0.59  $&$   998    _{-139 }^{+  477}$ & $ 1504 \pm   54$ & 1101 \\
			\smallskip
			HD~091824 &   F &          O7V  & $-4.90   _{-0.21} ^{+ 0.21 }$&  8.15 &  0.25  &  0.77  &$ 1.76\pm  0.32  $&$  2863    _{-266 }^{+  293}$ & $ 1817 \pm  194$ & 1817 \\
			\smallskip
			HD~092044 &   F &         B1Ib  & $-5.95   _{-0.46} ^{+ 0.59 }$&  8.31 &  0.43  &  1.42  &$ 2.44\pm  0.58  $&$  3702    _{-713 }^{+ 1153}$ & $ 2311 \pm  248$ & 2311 \\
			\smallskip
			HD~093222 &   G &        O6III  & $-5.90   _{-0.20} ^{+ 0.05 }$&  8.10 &  0.36  &  1.76  &$ 2.13\pm  0.28  $&$  2812    _{-243 }^{+   66}$ & $ 2368 \pm  270$ & 2368 \\
			\smallskip
			HD~101008 &   F &        B0III  & $-5.00   _{-0.30} ^{+ 0.42 }$&  9.16 &  0.27  &  0.93  &$ 2.28\pm  0.40  $&$  4434    _{-567 }^{+  951}$ & $ 2378 \pm  192$ & 2378 \\
			\smallskip
			HD~108927 &   F &          B5V  & $-1.21   _{-0.09} ^{+ 0.12 }$&  7.77 &  0.24  &  0.74  &$ 1.36\pm  0.12  $&$   444    _{ -18 }^{+   25}$ & $  335 \pm    7$ &  334 \\
			\smallskip
			HD~110336 &   F &          B8V  & $-0.49   _{-0.14} ^{+ 0.20 }$&  8.64 &  0.45  &  1.21  &$ 1.69\pm  0.18  $&$   384    _{ -25 }^{+   38}$ & $  309 \pm    3$ &  309 \\
			\smallskip
			HD~110946 &   F &        B2III  & $-3.55   _{-0.29} ^{+ 0.60 }$&  9.18 &  0.50  &  1.60  &$ 1.46\pm  0.46  $&$  1683    _{-207 }^{+  537}$ & $ 2048 \pm   95$ & 1798 \\
			\smallskip
			HD~112607 &   F &        B5III  & $-1.48   _{-0.29} ^{+ 0.42 }$&  8.10 &  0.31  &  0.85  &$ 0.77\pm  0.36  $&$   557    _{ -69 }^{+  120}$ & $  589 \pm   15$ &  577 \\
			\smallskip
			HD~112954 &   F &        B9III  & $-0.77   _{-0.50} ^{+ 0.27 }$&  8.39 &  0.57  &  1.75  &$ 1.61\pm  0.39  $&$   304    _{ -63 }^{+   40}$ & $  381 \pm    5$ &  324 \\
			\smallskip
			HD~129557 &   V &          B1V  & $-3.04   _{-0.53} ^{+ 0.50 }$&  6.09 &  0.23  &  0.53  &$ 1.07\pm  0.52  $&$   525    _{-113 }^{+  136}$ & $  409 \pm   17$ &  409 \\
			\smallskip
			HD~146284 &   F &        B9III  & $-0.77   _{-0.50} ^{+ 0.27 }$&  6.71 &  0.25  &  0.77  &$ 0.95\pm  0.39  $&$   219    _{ -45 }^{+   29}$ & $  202 \pm    3$ &  202 \\
			\smallskip
			HD~146285 &   F &         B9IV  & $+0.10   _{-0.26} ^{+ 0.22 }$&  7.93 &  0.32  &  1.23  &$ 1.88\pm  0.25  $&$   209    _{ -24 }^{+   23}$ & $  155 \pm    2$ &  155 \\
			\smallskip
			HD~147196 &   F &          B8V  & $-0.49   _{-0.14} ^{+ 0.20 }$&  7.04 &  0.27  &  0.84  &$ 1.84\pm  0.18  $&$   218    _{ -14 }^{+   21}$ & $  137 \pm    1$ &  137 \\
			\smallskip
			HD~148594 &   F &          B7V  & $-0.67   _{-0.15} ^{+ 0.15 }$&  6.89 &  0.21  &  0.65  &$ 1.26\pm  0.16  $&$   241    _{ -16 }^{+   17}$ & $  182 \pm    2$ &  182 \\
			\smallskip
			HD~152245 &   V &      B0.5III  & $-4.80   _{-0.32} ^{+ 0.45 }$&  8.39 &  0.36  &  1.08  &$ 1.96\pm  0.45  $&$  2639    _{-358 }^{+  600}$ & $ 1758 \pm  192$ & 1758 \\
			\smallskip
			HD~152249 &   G &         O9Ia  & $-7.00   _{-0.46} ^{+ 0.46 }$&  6.38 &  0.47  &  1.58  &$ 2.54\pm  0.50  $&$  2288    _{-434 }^{+  535}$ & $ 1469 \pm  131$ & 1469 \\
			\smallskip
			HD~170634 &   F &          B8V  & $-0.49   _{-0.14} ^{+ 0.20 }$&  9.85 &  0.69  &  2.05  &$ 2.16\pm  0.18  $&$   454    _{ -29 }^{+   45}$ & $  432 \pm    6$ &  432 \\
			\smallskip
			HD~170740 &   F &          B2V  & $-2.28   _{-0.46} ^{+ 0.46 }$&  5.75 &  0.47  &  1.37  &$ 1.28\pm  0.47  $&$   215    _{ -41 }^{+   50}$ & $  224 \pm   12$ &  224 \\
			\smallskip
			HD~185418 &   G &        B0.5V  & $-3.55   _{-0.36} ^{+ 0.42 }$&  7.49 &  0.52  &  1.39  &$ 1.84\pm  0.40  $&$   850    _{-131 }^{+  180}$ & $  692 \pm   24$ &  692 \\
			\smallskip
			HD~287150 &   F &        A1III  & $+0.73   _{-0.36} ^{+ 1.06 }$&  9.26 &  0.37  &  1.22  &$ 1.11\pm  0.71  $&$   289    _{ -44 }^{+  183}$ & $  399 \pm    5$ &  304 \\
			\smallskip
			HD~294304 &   F &          B6V  & $-0.89   _{-0.15} ^{+ 0.14 }$& 10.05 &  0.42  &  1.23  &$ 1.59\pm  0.16  $&$   874    _{ -59 }^{+   58}$ & $  741 \pm   18$ &  741 \\
			\smallskip
			HD~303308 &   V &          O6V  & $-5.20   _{-0.19} ^{+ 0.19 }$&  8.12 &  0.45  &  1.36  &$ 1.61\pm  0.29  $&$  2469    _{-202 }^{+  220}$ & $ 2196 \pm  217$ & 2196 \\
			\smallskip
			HD~315021 &   F &          B0V  & $-3.85   _{-0.34} ^{+ 0.34 }$&  8.57 &  0.34  &  1.24  &$ 1.73\pm  0.40  $&$  1719    _{-246 }^{+  287}$ & $ 1372 \pm  136$ & 1372 \\
			\smallskip
			HD~315023 &   F &          B2V  & $-2.28   _{-0.46} ^{+ 0.46 }$& 10.03 &  0.36  &  1.48  &$ 1.83\pm  0.46  $&$  1464    _{-277 }^{+  342}$ & $ 1246 \pm   46$ & 1246 \\
			\smallskip
			HD~315024 &   F &          B1V  & $-3.04   _{-0.53} ^{+ 0.50 }$&  9.63 &  0.30  &  1.20  &$ 2.11\pm  0.53  $&$  1966    _{-425 }^{+  510}$ & $ 1295 \pm   73$ & 1295 \\
			\smallskip
			HD~315032 &   F &          B1V  & $-3.04   _{-0.53} ^{+ 0.50 }$&  9.18 &  0.28  &  1.01  &$ 1.62\pm  0.54  $&$  1747    _{-378 }^{+  453}$ & $ 1321 \pm   89$ & 1321 \\
			
			\hline
		\end{tabular}
	\end{center}
	{ {  { Notes:}} Reddening curves (column~2) by F
          \citep{FM07}, G \citep{G09}, V \citep{V04}; SpL (column~3)
          by \cite{1} with associated $M_{V}$ (column~4) by
          \cite{40,41}; $V$-band (column~5) by \cite{Kharchenko09};
          luminosity distance $D_{{L}}(A^{\rm{Ref}}_{V})$ (column~9)
          using extrapolated $A^{\rm{Ref}}_V$ (column~7) estimated by
          the same authors of the reddening curves;  luminosity distance 
         $D_{{L}}(A_{V})$ (column~11) utilizing our estimate of $A_{V}$
          (column~8).}
\end{table*}
To account for an offset between both catalogues we introduce
$\sigma_C$; thus, $\sigma^2(M_V) = \sigma^2_{\rm {SpT}} +
\sigma^2_{\rm{LC}} + \sigma^2_{\rm C}$. For estimating the error in
the luminosity distance we add the photometric error
$\sigma(V)$. Fig.~\ref{DL2DGorg.pdf} shows the distance ratios $D_{\rm
  {L}}(A^{\rm {Ref}}_V) /D_{\rm {Gaia}}$ vs. $D_{\rm {Gaia}}$ for our
sample.  For the same star, the luminosity distance $D_{\rm
  {L}}(A^{\rm {Ref}}_V)$ generally overpredicts the Gaia distance
$D_{\rm {Gaia}}$.  There are no stars with a distance ratio below
$D_{{L}}(A^{\rm {Ref}}_V) /D_{\rm {Gaia}} < 1.25$ detected at $3
\sigma$ confidence. The distance ratios range between $0.7 \simless
D_{{L}}(A^{\rm {Ref}}_V) /D_{\rm {Gaia}} \simless 1.9$. The bottom
and top quartiles are at 1.09 and 1.37, which underlines the
overprediction in $D_{{L}}(A^{\rm {Ref}}_V)$. A dependency of
$D_{{L}}(A^{\rm {Ref}}_V)$ on the spectral types and luminosity
classes is not observed. We emphasize that stars of identical spectral
types show deviations in opposite directions which excludes any
systematic offsets in $M_V$ (Table~\ref{Tab1}). Therefore, we
interpret this discrepancy as to arise from an underestimate of
$A^{\rm {Ref}}_V$. Recent investigations \citep{46} were still not
able to detect any discrepancy within the errors of the Gaia data
release two \citep{14}; this could only be brought to light utilizing
the unprecedented resolution of the Gaia data release three.

\section{The dust model including submicrometer-sized grains \label{dust.sec}}

Our study utilizes a dust model{\footnote{Custom software
  AbsreDgaia available at https://github.com/rsiebenm/dark\_dust}
  by \cite{S23}} that aligns with current observational constraints of
dust in the diffuse ISM of the Milky Way \citep{HD21}.  It accounts
for representative solid phase element abundances of the main
absorbing dust components \deleted{(Sect.~\ref{abu.sec})} and explains
accurately phenomena such as wavelength-dependent reddening,
extinction, starlight polarization, and the emission of unpolarized
and polarized light seen by Planck \citep{Planck20}. Recently, the
(sub)millimeter excess emission in the Milky Way has gained an
alternative explanation by adjusting the grain emissivity at these
wavelengths \citep{HD21} to avoid micrometer-sized cold dust
particles. However, these models fail short resolving the distance
discrepancy reported in Sect.~\ref{disc.sec}.

The dust model incorporates three populations: 1) nanoparticles of
graphite, silicate, and polycyclic aromatic hydrocarbon (PAH), 2)
spheroidal grains of amorphous carbon and silicate at mean radii of
$\sim 30$\,nm, using the latest optical constants for amorphous
silicates \citep{Demyk22}, and 3) submicrometer-sized dust
particles\deleted{; the latter component has been labelled by
  \cite{10} as dark dust (DD)}.

Flat infrared extinction curves can also be explained by very porous,
fluffy particles \citep{KS94} with the degree of porosity constrained
through a comparison of optical and submillimeter polarization
\citep{Guillet18}. Submicrometer sized grains have been studied by
\cite{Mathis77, Vos4, 44} and have been included in other dust models
by \cite{Ormel11, Ysard24}. Recently, the impact of gray extinction on
type Ia supernovae distance measurements has been discussed by the
Dark Energy Survey Collaboration \citep{Popovic24}. Furthermore,
micrometer-sized particles from the diffuse ISM were also measured in
situ from the Ulysses, Galileo, and Stardust space probes \citep{26,
  27, 28}. They appear in sightlines connected to the cold ISM
\citep{10}.

\replaced{micrometer-sized grains}{Submicrometer-sized grains} absorb
a fraction of the interstellar radiation field, ISRF
\citep{29}. Because these grains are large, they are cold and emit at
long wavelengths. Originally very cold (10\,K) dust emission was
detected in our Galaxy towards high-density regions \citep{30} and in
non-active galaxies \citep{31}. Such cold dust was confirmed by ISO
\citep{32,33}. More recently, excess emission at 0.5\,mm observed by
Herschel cannot be explained by a single modified black-body
temperature component \citep{34, 35,36}; these results were confirmed
with ALMA \citep{37} and LABOCA \citep{38} at even longer wavelengths.

\section{The distance including submicrometer-sized grains}
Correcting the distance discrepancy is not feasible through arbitrary
adjustments to commonly used literature values in $A^{\rm{Ref}}_V =
R_V / E(B-V)$ and $M_{V}$ inserted in Eq.~\ref{Eq.1}. Several stars
with identical SpL show different distance ratios and thus would
require different adjustments in $M_V$ to reach $D_{{L}} = D_{\rm
  {Gaia}}$. The uncertainties in $M_{V}$ range from 0.25 to
0.5\,mag. However, for stars with $D_{{L}} / D_{\rm {Gaia}}
\simgreat 1.2$, modifications of $M_V > 0.5$\,mag, -- typically $0.8
\pm 0.3$\,mag -- would be necessary to align $D_{{L}}$ with $D_{\rm
  {Gaia}}$. Similarly, the typical error in $E(B-V)$ is around
0.03\,mag; for stars with $D_{{L}} / D_{\rm {Gaia}} \simgreat 1.2$,
modifications of $0.14 \simless E(B-V) \simless 0.56$\,mag would be
required to achieve distance unification. Consequently, we retain
$E(B-V)$ and substitute the extrapolated $R_V$ parameter with our dust
model, utilizing Eqs.~\ref{Eq.3}~-~\ref{Eq.4}.


\begin{figure} [!htb]
	\begin{center}
	  \includegraphics[width=9cm,clip=true,trim=4.2cm 5.5cm 3.cm 5.6cm]{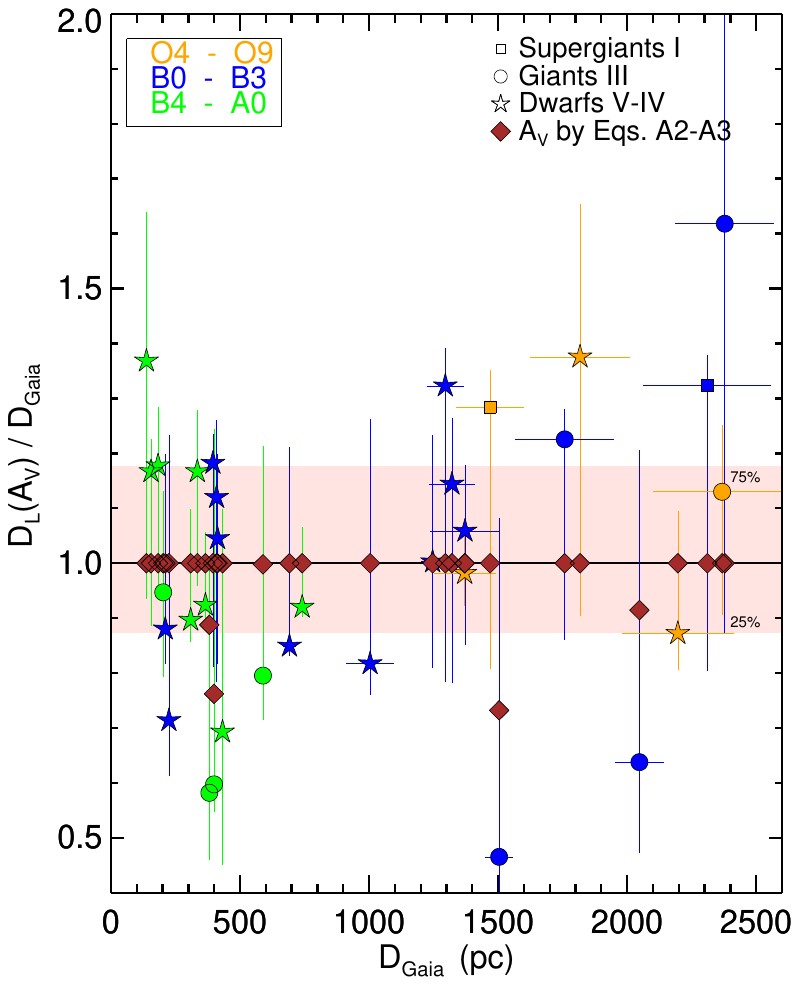}
	\end{center}
	\caption{Corrected distance ratios $D_{{L}}(A_{V}) /D_{\rm
            {Gaia}}$ vs. $D_{\rm {Gaia}}$. {  {The luminosity distance
          derived with $A_V$ using Eqs.~\ref{Eq.3}-\ref{Eq.4}
          (diamonds in brown) unifies the distance ratio for all but
          four stars. These four stars still agree within their errors
          with a ratio of one. For comparison, the luminosity distance
          determined from $A_V$ is derived by a single-parameter model
          fit to the reddening in the $UBV JHK$ bands, assuming a mass
          fraction of submicrometer particles of 0.55\,g in a total of
          1\,g of dust (symbols in color, as denoted in
          Fig.~\ref{DL2DGorg.pdf}). The shaded area depicts the top
          and bottom quartiles of the single-parameter model
          distribution, with a mean of one (solid line). }} }
	\label{DL2DGcor.pdf}
\end{figure}

Fig.~\ref{DL2DGcor.pdf} shows the corrected distance ratios
$D_{{L}}(A_{V})/D_{\rm {Gaia}}$ vs. $D_{\rm {Gaia}}$ for our sample,
where $D_{{L}}(A_{V})$ denotes the luminosity distance derived from
$A_V$ using Eq.~\ref{Eq.1}.  The new luminosity distances
$D_{{L}}(A_{V})$, agree with $D_{\rm {Gaia}}$ within $0.75 <
D_{{L}}(A_{V}) / D_{\rm {Gaia}} \simless 1$ and show a $1 \sigma$
scatter of 6\,\% around a median of one.  The four stars with
$D_{{L}}(A_{V}) > D_{\rm {Gaia}}$ still agree within their errors with
equal distances $D_{{L}}(A_{V}) = D_{\rm {Gaia}}$. For 29 out of 33
stars both distances agree within 2\%. Their scatter of the distance
ratio is reduced by a factor of five when compared to the literature
values $D_{{L}}(A_{V}^{\rm {Ref}})$, which are, for ease of comparison,
also shown in Fig.~\ref{DL2DGcor.pdf}.

The reliability of the $A_{V}$ estimate (Eq.~\ref{Eq.1}) is validated
through fitting the absolute reddening curve across the entire
observed wavelength range using the three-component dust model, which
incorporates \replaced{micrometer-sized grains}{\dd}. {  {Models
    that incorporate a simple continuation of the dust size
    distribution to micrometers for the amorphous carbon and silicate
    grains or ignore submicrometer-sized grains fail to align to
    explain the reddening at $\lambda > 1\mu$m and often violate
    abundance constraints (Table~\ref{Tab2DD}).}}

The submicrometer-sized dust population typically contributes to
one-third of the total extinction $A_V$. This dust component
constitutes $56 \pm 12$\, \% of the total dust mass and depends on the
environmental conditions and chemical composition of the clouds along
the sightlines. Note that the total gas-to-dust mass ratio remains
unchanged in the models $M_{\rm {gas}}/M_{\rm {dust}}$
(Table~\ref{abu.tab}). {  {The addition of submicrometer-sized
    grains only affects the relative mass distribution of different
    dust particle types within 1\,g of dust (Eq.~\ref{Eq.5}). It does
    not change the total dust mass or the amount of elements depleted
    from the gas phase (Table~\ref{Tab2DD}). Models that ignore the
    submicrometer-sized grain component or merely extend the size
    distribution of amorphous particle components from 250\,nm to a few
    microns often fail to explain the reddening at $\lambda > 1\mu$m
    and violate abundance constraints. This is evident when comparing
    the goodness-of-fit to the IR data for models with,
    $\chi^{2}_{\mathrm{IR}}$, and without,
    $\chi^{2,\text{no-s$\mu$}}_{\mathrm{IR}}$, submicrometer-sized
    grains, as $\chi^{2}_{\mathrm{IR}} <
    \chi^{2,\text{no-s$\mu$}}_{\mathrm{IR}}$ (Table~\ref{Tab2DD}). }}

Without knowledge of the trigonometric distance, one can assume that
half of the dust mass resides in submicrometer grains to account for
the offset in the luminosity distance reported in
Fig.~\ref{DL2DGorg.pdf}. { {This is demonstrated in
    Fig.~\ref{DL2DGcor.pdf} using a single-parameter model
    {\footnote{Custom software 1\_absReddEbv available at
      https://github.com/rsiebenm/dark\_dust}} to fit the reddening in
    the $U B V J H K$ bands. The visual extinction $A_V$ is derived
    from the reddening at infinite wavelengths based on the best-fit
    model, with the carbon abundance in amorphous grains treated as
    the only free parameter. }}

\section{Conclusion}

We derived the visual extinction through the diffuse ISM from a
comparison of the luminosity distance and the trigonometric distance
of 33 nearby ($\lesssim 2.5$\,kpc) early-type stars using the most
comprehensive sample of reddening curves and SpL identifications
currently accessible. {  {Our analysis supports the presence of a
    submicrometer-sized dust population in 80\% of the sightlines,
    consistent with the dust model introduced by \cite{Mathis77} and
    employed to explain IR reddening by \cite{WL15a}.}} Due to the
wavelength-independent extinction of such very large grains such a
population is hidden in photometric data from the optical/far UV
range. Dust models ignoring the submicrometer-sized grain component
thus underestimate the total extinction of the sightline and -- as a
consequence -- resulting in significantly larger distance estimates
when applying traditional photometric distance measurements.

Apart from the importance of the gray extinction component, our study
has developed a consistent dust model that satisfies contemporary
constraints on dust in the general ISM field \citep{HD21} and aligns
with the available data for individual sightlines in our sample. { 
  {We also provide a single-parameter model to fit the optical/IR
    reddening and correct the average overprediction in the luminosity
    distance for the sample, even when assuming the parallax is
    unknown.}}

The dust model respects estimates of the visual extinction derived
from the Gaia parallax. Our study emphasizes the significance of
considering the absolute reddening of individual sightlines across the
entire wavelength range and in particular in the IR instead of
utilizing the extrapolated parameter $R_V$. 

{  {Besides our own ongoing efforts to increase the sample size and
    enhance statistical robustness, we hope to inspire the community
    to further test the model using the custom software provided.}} To
better constrain the population of \dd \/ in the diffuse ISM, targeted
observations should focus on stars with known spectral types,
luminosity classes, and precise trigonometric distances provided by
Gaia or VLTI. For these stars, the depletion of elements such as C,
Si, Mg, Fe, O, and Al from the gas phase should be measured. These
elements contribute significantly to the dust extinction. For these
sightlines, the reddening $E(\lambda - V)$ should be well established
in the far UV, around the 2175\,\AA\, bump, and in the NIR/MIR.  The
latter can be complemented by observations with the JWST, which offers
spectral information that can potentially detect or provide stringent
upper limits on ice absorption bands in the diffuse ISM. The silicate
stoichiometry can be revealed through MIR spectro-polarimetry
\citep{Wright02}. The geometry and particle shape of \dd \/ can be
determined by combining high spatial resolution observations of
polarized emission in the millimeter continuum, as provided by BLAST
\citep{Blast}, with dichroic polarization of starlight in the optical
\citep{Bagnulo17}. Such studies of submicrometer-sized grains appear
also of interest for dark energy surveys, as analyzed by
\cite{Popovic24}.


\begin{acknowledgements}
	{We are grateful to Tereza Jerabkova and Miguel Vioque for 
		discussions of Gaia distances.}
\end{acknowledgements}


\bibliography{apj_letter_distance}{}
\bibliographystyle{aasjournal}

\appendix

\section{Absolute reddening}

\noindent
The amount of visual extinction, denoted by $A_V$, which is necessary
to align $D_{{L}}$ with $D_{\rm {Gaia}}$ is calculated using the
photometric equation (Eq.~\ref{Eq.1}) with errors {  {including
  $G$-band brightness corrections by \cite{MA22} }}

\begin{equation} \label{Eq.2err}
	\sigma(A_V)^2 = \sigma(V)^2 + \sigma(M_V)^2 + \left( {{5}
		\over {\log 10}}  \, {{\sigma(\pi,G)} \over {D_{\rm {Gaia}}}} \right)^2
\end{equation}
\deleted{
In Fig.~\ref{dAv.pdf} we compare the difference between our estimate
of the visual extinction $A_V$ (Eq.~\ref{Eq.1}) and the estimate
$A^{\rm{Ref}}_V$, which was obtained by extrapolation of the reddening
curves \citep{V04, FM07, G09}. Similar to Fig.~\ref{DL2DGorg.pdf} and
irrespective of the SpL, the latter estimate $A^{\rm{Ref}}_V$ exhibits
a general underestimate of about half a magnitude when compared to
$A_V$.
}

The optical depth $\tau_V = A_V / 1.086$ is related to the column
densities of nanoparticles and the amorphous carbon and silicate particles $N_n$ and
to the column density of submicrometer-sized grains $N_{s\mu}$; the
corresponding mass extinction cross-section $K_n$ and
$K_{s\mu}$\,(g/cm$^3$) are based on the dust model \citep{S23}:

\begin{equation} \label{Eq.3}
	\tau_V = N^n \ K^n_V \ + \ N^{s\mu} \ K^{s\mu}_V   \ .		
\end{equation}

The extinction cross-section \citep{42, Vos4, 44} diminishes at
infinite wavelengths, i.e. $K(\infty) = 0$. To prevent negative
optical depths, we assume that the reddening at infinite wavelengths
is smaller than in the $K$-band, hence $A_V = -E(\infty) > -E(K)$ { 
  {, and ignore models not respecting this conditon}}. The reddening
$E(B-V) = 1.086 \ (\tau_B - \tau_V)$ provides a second constraint:

\begin{equation} \label{Eq.4}
	\tau_B - \tau_V = N^n \ (K^n_B - K^n_V) \ + \ N^{s\mu} \  (K^{s\mu}_B  -  K^{s\mu}_V) \ .
\end{equation}

These two equations enable us to derive the {  {relative}} mass of
each component, specifically $m_n = N_n / (N_n + N_{s\mu})$ of the
nano- and amorphous grains and $m_{s\mu} = N_{s\mu} / (N_n +
N_{s\mu})$ of the submicrometer-sized particles. Notably, our approach
to computing $A_V$ is validated through a fit to the observed
reddening curve using our dust model and obviates the need for the
extrapolated parameter $R_V = A_V / E(B-V)$.

\section{Fitting reddening curves \label{fit.sec}}

The methodology for calculating the wavelength-dependent extinction
cross-section $K(\lambda)$ for partially aligned and wobbling
spheroidal grains, nanoparticles, and PAH, from the optical constants
of dust materials is outlined in \cite{S23}. The normalized reddening
curves $E(\lambda - V) / E(B-V)$ for the sample are observed
spectroscopically between $0.09 - 0.27\mu$m with IUE/FUSE satellites
and in the $UBV$- and $JHK$-bands with references listed in column~2
of Table~\ref{Tab1}. These curves are transformed into absolute
reddening that are used in this work by multiplying them with the
corresponding reddening $E(B-V)$ provided in those same
references. The extrapolated reddening at infinitely long wavelengths
is substituted with the visual extinction $A_V$, derived from
Eq.~\ref{Eq.1}, and are specified in Table~\ref{Tab1}. By adjusting
grain sizes and abundances within the three populations, we achieve
the best fit for the absolute reddening curve of each star, surpassing
previous models that solely addressed relative reddening or extinction
curves.

{ {To achieve the optimal fits for the reddening curves, a set of
    seven best-fit parameters was computed by a least
    $\chi^2$-technique using the Levenberg--Marquardt algorithm as
    implemented in MPFIT}}\footnote{http://purl.com/net/mpfit}
\citep{Markwardt09}. { {The distribution of the various grain
    components in 1\,g of dust is treated as a set of free parameters,
    defined by the relative mass of}} the submicrometer-sized grains
($m_{s\mu}$), the amorphous silicates ($m_{\rm{Si}}$) and carbon
($m_{{aC}}$), and the nanoparticles of silicates ($m_{\rm{vSi}}$),
graphite ($m_{\rm{vgr}}$), and PAH ($m_{\rm{PAH}}$). {  {The { 
    {relative}} mass of component $i$ in 1\,g of dust, is computed
  from }}

\begin{equation} \label{Eq.5}
	m_{\rm {i}} = {\frac {\displaystyle{ \ \ \ \/ \ \ \ \mu_{\rm i}  \  \frac{[\rm{X_{\rm {i}}}]}{[\rm{H}]}}}
		{\displaystyle{{\sum_{i}{\ \mu_{\rm i}  \ {\frac{[\rm{X_{\rm {i}}}]}{[\rm{H}]}}}}}}}\ \,,
\end{equation}
\noindent
where the relative dust abundances of an element, which is in our
models either C or Si, with respect to H are denoted by $[\rm{X_{\rm
      {i}}}]/[\rm{H}]$ and the molecular weights $\mu_i$ as specified
in Table~\ref{abu.tab}.

{  {Models that violate the depletion constraints for C and Si
    atoms, specifically C/Si $< 5.2$ \citep{HD21,S23} are
    excluded. The total dust mass $M_{\rm {dust}}$ is estimated by
    summing all atoms depleted from the gas phase and scaling by the
    molecular weights corresponding to the assumed grain
    stoichiometry. The gas mass $M_{\rm {gas}} \sim 1.4 \ M_{\rm {H}}$
    is calculated by summing the contributions of helium and hydrogen,
    assuming a He:H ratio of 1:10. The derived gas-to-dust mass ratio,
    $M_{\rm {gas}} / M_{\rm dust} \sim 125$ \citep{HD21}, remains
    approximately constant (Table~\ref{abu.tab}). Consequently, the
    relative mass distribution in 1\,g of dust in our models aligns
    with the material available in the ISM for grain formation.}}

{  {A power-law dust size distribution is characterized by the
    exponent $q$ and the lower and upper radii. For amorphous carbon
    and silicate grains, we use a minimum radius of 6\,nm. Following
    \cite{Mathis77}, we set the relatively unconstrained upper radii
    to fixed values, with $r^+_{\rm{Si}} = r^+_{{aC}} =250$\,nm, which
    also serves as the minimum radius for the submicrometer grains,
    and $r^+_{\rm{s}\mu}=3 \mu$m. The}} mass extinction cross sections
$K_{i}(r)$\, (cm$^2$/g-dust) of a dust particle of population $i$ with
volume averaged radius $r$ and bulk density $\rho_i$\, is

\begin{equation} \label{Eq.6}
	K_{i}(r) = \frac {m_i} {{\displaystyle {\frac {4 \pi}{3}}
			\ \rho_i}} \ \frac {r^{-q}} { \displaystyle
		\int_{r_{-,i}}^{r_{+,i}} r^{3-q} \ \mathrm{d}r} \ C_{i}(r) \ .
\end{equation}

The cross sections $C$ are derived using efficiency factors $Q$, and
are computed for spheroids using \cite{V93} and custom software
provided by \cite{Vos4}.  Fine-tuning of the fits to the 2175\,\AA \,
bump involves allowing the central wavelength ($x_0$) and damping
constant ($\gamma$) of the Lorentzian profiles of PAH absorption
cross-section to remain free. This leads to in total seven adjustable
parameters which are treated in the minimization algorithm.

\begin{figure*} [!htb]
	\begin{center}
		\includegraphics[width=18.4cm,clip=true,trim=1.5cm 3.5cm 1cm 4cm]{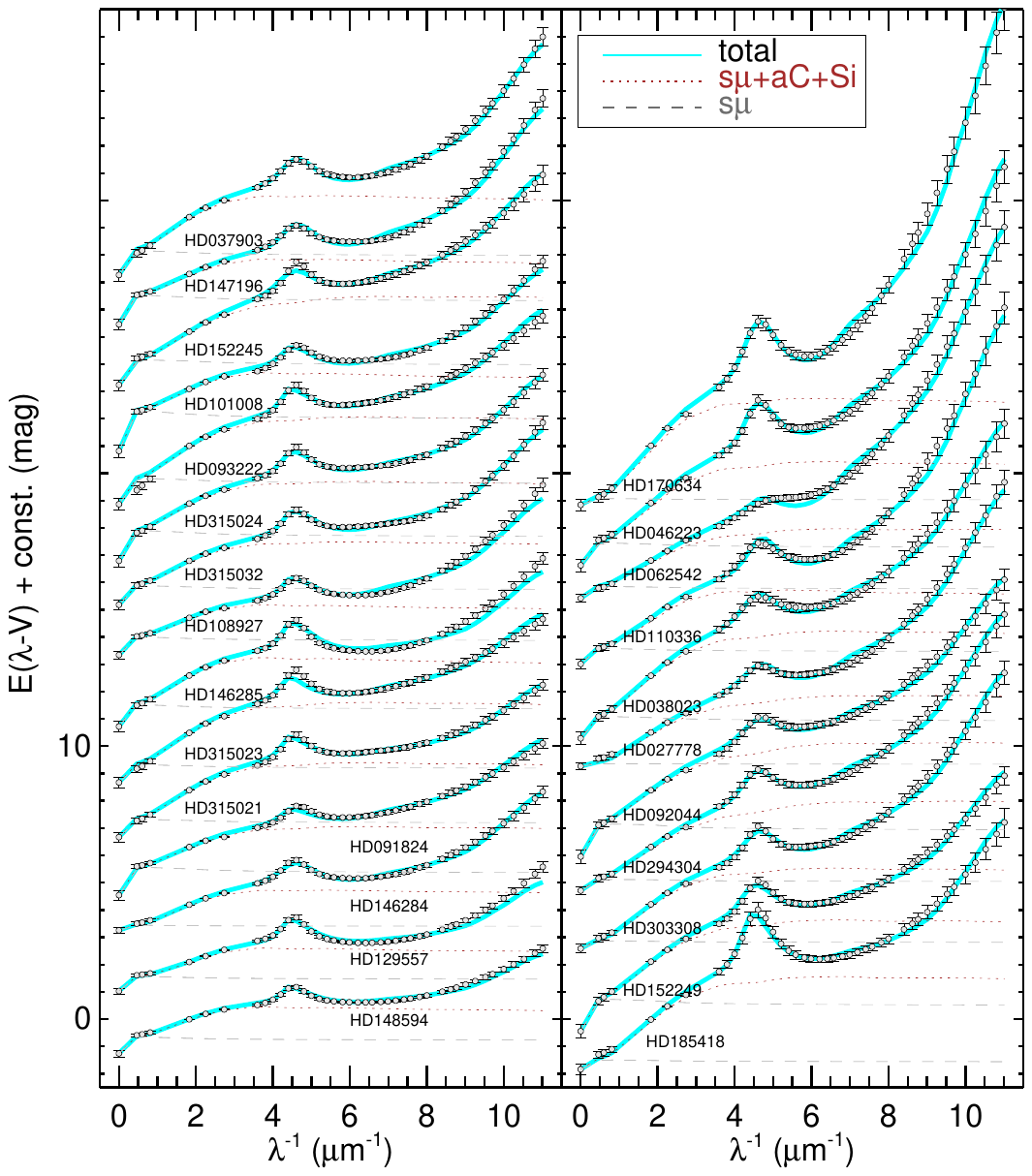}
	\end{center}
	\caption{The absolute reddening curves $E(\lambda - V)$ of the
          sample, shifted by an offset for better visibility. Data
          points (circles) cover the range $0.09 - 2.2\,\mu$m and are
          complemented at infinite wavelengths by $-A_V$
          (Table~\ref{Tab1}). The best fit (cyan) with contributions
          from the amorphous carbon and silicates (brown), and
          submicrometer-sized grains (gray), are shown, with model
          parameters listed in Table~\ref{Tab2DD}. Notable is the
          dominance of the submicrometer-sized grains in the infrared
          and their wavelength-independent contribution to the
          optical/far UV, respectively.}
	\label{Fig3.pdf}
\end{figure*}
\newpage
\begin{table*}[!htb]
	\begin{center}  
		\caption {Dust abundances (ppm), gas-to-dust mass
                  ratios, molecular weights and bulk densities { 
                    {of the best-fit.}} \label{abu.tab}}
		\begin{tabular}{c | r r r r  r | r r r  r | c}
			\hline\hline
			Star     & 
			$\frac{[C]}{[H]}_{\rm {tot}}$ &
			$\frac{[C]}{[H]}_{\rm {s\mu}}$ &
			$\frac{[C]}{[H]}_{\rm {aC}}$ &
			$\frac{[C]}{[H]}_{\rm {vgr}}$ &
			$\frac{[C]}{[H]}_{\rm {PAH}}$ &
			$\frac{\rm{[Si]}}{\rm{[H]}}_{\rm {tot}}$ &
			$\frac{\rm{[Si]}}{\rm{[H]}}_{\rm {s\mu}}$ &
			$\frac{\rm{[Si]}}{\rm{[H]}}_{\rm {aSi}}$ &
			$\frac{\rm{[Si]}}{\rm{[H]}}_{\rm {vSi}}$ &
			$\frac{M_{\rm {gas}}}{M_{\rm {dust}}}$ \\
			\hline
  HD~027778 &    95 &    23 &    48 &    15 &     9 &    37 &     8 &    18 &    11 &   125 \\
  HD~037903 &   136 &    48 &    63 &    19 &     6 &    44 &    16 &    20 &     9 &   116 \\
  HD~038023 &    91 &    36 &    37 &    12 &     5 &    39 &    12 &    19 &     8 &   125 \\
  HD~046223 &    89 &    40 &    25 &    20 &     4 &    44 &    13 &    20 &    11 &   123 \\
  HD~062542 &    86 &    41 &    21 &    13 &    11 &    52 &    13 &    20 &    18 &   120 \\
  HD~091824 &    89 &    42 &    34 &     9 &     4 &    39 &    14 &    19 &     6 &   125 \\
  HD~092044 &    90 &    39 &    36 &    11 &     5 &    39 &    13 &    19 &     7 &   125 \\
  HD~093222 &   103 &    37 &    51 &    11 &     4 &    34 &    12 &    17 &     5 &   125 \\
  HD~101008 &   103 &    48 &    37 &    13 &     5 &    43 &    16 &    20 &     7 &   121 \\
  HD~108927 &   100 &    45 &    32 &    19 &     4 &    43 &    15 &    20 &     8 &   122 \\
  HD~110336 &    78 &    39 &    14 &    23 &     2 &    45 &    13 &    20 &    12 &   124 \\
  HD~129557 &    76 &    41 &     8 &    19 &     8 &    42 &    13 &    24 &     4 &   126 \\
  HD~146284 &    97 &    32 &    40 &    16 &     9 &    37 &    10 &    19 &     7 &   125 \\
  HD~146285 &   106 &    37 &    48 &    17 &     5 &    33 &    12 &    18 &     4 &   125 \\
  HD~147196 &   108 &    50 &    30 &    21 &     6 &    47 &    16 &    20 &    10 &   119 \\
  HD~148594 &    97 &    39 &    40 &    13 &     5 &    36 &    13 &    20 &     4 &   125 \\
  HD~152245 &    93 &    38 &    37 &    10 &     8 &    38 &    12 &    19 &     6 &   125 \\
  HD~152249 &    94 &    38 &    38 &    10 &     7 &    38 &    12 &    20 &     6 &   125 \\
  HD~170634 &    85 &    27 &    31 &    22 &     5 &    41 &     9 &    20 &    12 &   125 \\
  HD~185418 &    98 &    29 &    41 &    17 &    11 &    34 &    10 &    20 &     5 &   125 \\
  HD~294304 &    92 &    32 &    39 &    14 &     6 &    38 &    10 &    19 &     9 &   125 \\
  HD~303308 &    93 &    31 &    42 &    15 &     6 &    38 &    10 &    19 &     9 &   125 \\
  HD~315021 &   102 &    36 &    51 &    10 &     5 &    35 &    12 &    19 &     4 &   125 \\
  HD~315023 &   103 &    35 &    55 &     7 &     7 &    34 &    11 &    18 &     5 &   125 \\
  HD~315024 &   103 &    38 &    47 &    13 &     5 &    34 &    12 &    16 &     5 &   125 \\
  HD~315032 &   129 &    45 &    57 &    22 &     5 &    42 &    15 &    20 &     8 &   118 \\
			\hline 
			$\mu$  &    12 & 12    &12     &12     &12     &     &   135 &   100 &   135 & \\
			$\rho$\, (g/cm$^{3}$)&  &1.8    &1.6    &2.2    &    \phantom{}     &       &  3.4 &   2.7 &   3.5 &       \\
			\hline
		\end {tabular}
	\end{center}
	{  {Notes:}} Assumed stoichiometry of silicate material in nano
	and submicrometer particles is the nominal of
	Mg$_{1.3}$(Fe,Ni)$_{0.3}$SiO$_{3.6}$ by \cite{DH21} and for
	amorphous silicates the 97:3 mix in mass composed of MgSiO$_3$ and
	Mg$_{0.8}$Fe$^{2+}_{0.2}$SiO$_3$ by \cite{Demyk22}.
\end {table*}

\newpage
\begin{table*}[!htb]
	\begin{center}  
		\caption {Model parameters for fits to the absolute reddening curve. \label{Tab2DD}}
		\begin{tabular}{c c c  r r c  c c  c c c}
			\hline\hline
			Star & $m_{s\mu}$&  $m_{\rm {Si}}$&  $m_{\rm{vSi}}$&  $m_{\rm{aC}}$&  $m_{\rm{vgr}}$&  $m_{\rm{PAH}}$& $q$ & $\chi^2$ & $\chi^2_{\rm {IR}}$ & $\chi^{2, \text{no-s$\mu$}}_{\mathrm{IR}}$ \\
			& (\%) &  (\%)      &   (\%)     &    (\%)    &    (\%)    &   (\%) &  & & \\
			\hline
  HD~027778 &    31 &    30 &    25 &     9 &     3 &     2 &   2.7  &   0.9  &   2.4  &    5.8  \\
  HD~037903 &    63 &    17 &    10 &     7 &     2 &     1 &   2.2  &   0.6  &   1.3  & 43.3$^*$  \\
  HD~038023 &    56 &    23 &    13 &     6 &     2 &     1 &   2.5  &   0.7  &   0.1  &   39.9  \\
  HD~046223 &    55 &    22 &    16 &     3 &     3 &     1 &   2.4  &   1.0  &   0.5  &   60.0  \\
  HD~062542 &    46 &    22 &    26 &     3 &     2 &     1 &   2.5  &   1.4  &   0.7  &   10.7  \\
  HD~091824 &    69 &    18 &     8 &     4 &     1 &     0 &   2.4  &   0.6  &   0.4  &151.8$^*$  \\
  HD~092044 &    62 &    21 &    10 &     5 &     1 &     1 &   2.7  &   0.6  &   0.1  & 92.4$^*$  \\
  HD~093222 &    64 &    20 &     8 &     7 &     1 &     1 &   2.4  &   1.4  &  11.6  & 16.6$^*$  \\
  HD~101008 &    74 &    14 &     7 &     3 &     1 &     0 &   2.1  &   0.9  &   0.2  &251.8$^*$  \\
  HD~108927 &    66 &    18 &    10 &     3 &     2 &     0 &   2.0  &   1.5  &   0.1  & 62.4$^*$  \\
  HD~110336 &    53 &    23 &    18 &     2 &     3 &     0 &   2.2  &   1.3  &   0.1  &   21.6  \\
  HD~129557 &    66 &    24 &     6 &     1 &     2 &     1 &   2.0  &   1.7  &   0.3  & 61.2$^*$  \\
  HD~146284 &    48 &    27 &    14 &     7 &     3 &     2 &   2.1  &   0.4  &   0.2  &  5.9$^*$  \\
  HD~146285 &    66 &    19 &     5 &     6 &     2 &     1 &   2.0  &   2.0  &   0.4  & 52.2$^*$  \\
  HD~147196 &    69 &    15 &    11 &     3 &     2 &     1 &   2.1  &   0.9  &   0.1  &142.5$^*$  \\
  HD~148594 &    68 &    19 &     5 &     5 &     2 &     1 &   2.0  &   0.6  &   0.3  & 70.1$^*$  \\
  HD~152245 &    62 &    22 &     9 &     5 &     1 &     1 &   2.6  &   0.7  &   0.6  & 85.8$^*$  \\
  HD~152249 &    63 &    22 &     8 &     5 &     1 &     1 &   2.5  &   0.6  &   0.2  & 87.8$^*$  \\
  HD~170634 &    36 &    30 &    24 &     6 &     4 &     1 &   2.3  &   1.3  &   0.1  &    2.5  \\
  HD~185418 &    47 &    31 &     9 &     7 &     3 &     2 &   2.5  &   0.9  &   2.1  &   18.6  \\
  HD~294304 &    47 &    27 &    16 &     7 &     2 &     1 &   2.7  &   0.6  &   0.1  &   11.7  \\
  HD~303308 &    44 &    27 &    17 &     7 &     3 &     1 &   2.5  &   0.7  &   0.3  &    7.6  \\
  HD~315021 &    61 &    23 &     6 &     7 &     1 &     1 &   2.4  &   0.4  &   0.5  & 28.2$^*$  \\
  HD~315023 &    59 &    23 &     8 &     8 &     1 &     1 &   2.4  &   1.0  &   2.2  & 15.8$^*$  \\
  HD~315024 &    68 &    17 &     7 &     6 &     2 &     1 &   2.4  &   0.4  &   0.7  & 95.3$^*$  \\
  HD~315032 &    63 &    18 &     9 &     6 &     2 &     0 &   2.3  &   0.4  &   0.1  &  49.7$^*$  \\
			\hline
			\end {tabular}
		\end{center}
	    {  {Notes:} The symbol  $^{*}$  indicates that abundance constraints are violated.}
\end {table*}
\end{document}